\documentclass[
    reprint,
    preprintnumbers,
    superscriptaddress,
    nofootinbib,
     amsmath,amssymb,
     aps,
     prd,
    floatfix,
    ]{revtex4-2}
    
\usepackage{bm}
\usepackage{graphicx}
\usepackage[caption=false]{subfig}
\usepackage[usenames,dvipsnames]{xcolor} 
\usepackage[utf8]{inputenc}
\usepackage{color}
\usepackage[normalem]{ulem} 
\usepackage{physics}
\usepackage{mathrsfs}
\usepackage[colorlinks]{hyperref}
\hypersetup{ 
	colorlinks=true, 
	linkcolor=black, 
	filecolor=black, 
	citecolor = blue,       
	urlcolor=blue, 
} 
\usepackage{cleveref}
\usepackage{tensor}
\newcommand{\beq}{\begin{equation}}
\newcommand{\eeq}{\end{equation}}

\newcommand{\lag}{\mathcal{L}}

\NewDocumentCommand{\evat}{sO{\bigg}mm}{%
  \IfBooleanTF{#1}
   {\mleft. #3 \mright|_{#4}}
   {#3#2|_{#4}}%
}

\begin{document}
\count\footins = 1000 

{\hfill MIT-CTP/6011}
\title{On curvature corrections for field theory cosmic strings}

\author{Josu C. Aurrekoetxea}	
\email{jaurreko@mit.edu}
\affiliation{Center for Theoretical Physics, Massachusetts Institute of Technology, Cambridge, MA 02139, USA}
\author{Jose J. Blanco-Pillado}
\email{josejuan.blanco@ehu.eus}
\affiliation{IKERBASQUE, Basque Foundation for Science, 48011, Bilbao, Spain,}
\affiliation{EHU Quantum Center, University of the Basque Country, UPV/EHU}
\affiliation{Department of Physics, University of the Basque Country, UPV/EHU, 48080, Bilbao, Spain}
\author{Alberto Garc\'ia Mart\'in-Caro}
\email{alberto.garcia.martin-caro@uvigo.gal}
\affiliation{EHU Quantum Center, University of the Basque Country, UPV/EHU}
\affiliation{Department of Physics, University of the Basque Country, UPV/EHU, 48080, Bilbao, Spain}
\affiliation{Instituto de Física, Computación e Ciencias Aeroespaciais (IFCAE), Universidade de Vigo. 32004 Ourense, Spain}
\author{J.M. Queiruga}
\email{xose.queiruga@usal.es}
\affiliation{Department of Applied Mathematics, University of Salamanca,
Casas del Parque 2, 37008 - Salamanca, Spain}
\affiliation{Institute of Fundamental Physics and Mathematics, University of Salamanca, Plaza de la Merced 1, 37008 - Salamanca, Spain}


\begin{abstract}

We present a combined analytical and numerical study of the effective action of field theory cosmic strings in the Abelian-Higgs model in flat space. Starting directly from the underlying solitonic field theory description, we provide a systematic derivation of the low energy effective action and present evidence for the absence of nontrivial curvature correction terms when only the translational Goldstone modes are retained. Using this framework, we extend the effective theory to include higher energy fluctuations of the soliton profile, which map to massive degrees of freedom propagating on the worldsheet. We show that the leading curvature contribution enters only through the coupling between these massive modes and the worldsheet Ricci scalar. We validate the resulting effective theory via lattice simulations of the full field theory equations of motion in flat space, implemented with Adaptive Mesh Refinement to capture the string dynamics across different scales. The numerical simulations confirm the dynamics obtained using the effective action in its validity range. Furthermore, they also demonstrate the existence of the predicted parametric instability of excited strings that drives the transfer of energy from massive excitations to the Goldstone sector.
\end{abstract}

\maketitle

\newpage

\section{Introduction}

Cosmic strings are relic topological defects expected to have formed during a phase transition in the early universe in certain extensions of the Standard Model \cite{Kibble:1976sj}. Their formation and evolution have been extensively studied over the past forty years using both analytical and numerical techniques \cite{Vilenkin:2000jqa}, with discrepancies between different approaches remaining a subject of debate (see, \cite{Blanco-Pillado:2023sap} and references therein). Given the large separation of scales between their thickness and length, the numerical study of such objects via lattice field theory simulations is computationally challenging. It is therefore natural to seek effective, low-energy descriptions such as the Nambu-Goto (NG) action, which captures the dynamics of an infinitely thin relativistic string \cite{nambu1970lectures,Goto:1971ce}. This action can be improved by incorporating curvature corrections that account for the finite thickness of the string in field theory. 

Various aspects of these curvature corrections have previously been examined in the literature, leading to somewhat differing results (see, for example, \cite{Maeda:1987pd,Gregory:1988qv,Anderson:1997ip}). In this work, we adopt the approach developed in Ref. \cite{Blanco-Pillado:2024bev}, where a systematic framework was established for deriving the effective action of domain walls directly from the underlying scalar field theory. In the general case, it was shown that corrections to the NG action appear as a finite set of terms known as the Dirac-Born-Infield (DBI) Galileons \cite{deRham:2010eu}. This is natural from the bottom-up effective field theory (EFT) interpretation, since a domain wall can be interpreted as a brane spontaneously breaking the Poincaré symmetry of the background spacetime.

In this work, we generalize this procedure to higher codimension defects, focusing on the Abelian-Higgs model as a field theory toy model that admits local string-like solutions in the form of Abrikosov-Nielsen-Olesen vortices \cite{Abrikosov:1957wnz,Nielsen:1973cs}. Such vortex strings represent the lowest-dimensional realization of higher codimension branes in the language of \cite{Hinterbichler:2010xn}. Hence, our results provide a “top-down” realization of the EFT construction of \cite{deRham:2010eu}. We show that no such corrections appear in the effective action for the zero modes of the string and that the leading deviations from NG dynamics in field theory strings arise from the excitation of higher energy degrees of freedom.

Building upon this foundation, we formulate a systematic framework for incorporating the massive excitations of the underlying field theory string into the NG action. Our analysis establishes that the Nambu-Goto action represents the unique contribution to the effective action describing string-like defects at the lowest order, in other words, provided that no additional high energy degrees of freedom are taken into account. At next-to-leading order, however, a new interaction arises between the first internal excitation mode of the string profile and the curvature of the worldsheet. We show how this interaction manifests itself as a direct coupling between the internal mode and the Ricci scalar constructed from the induced metric on the string worldsheet.

Our numerical simulations further indicate that this coupling gives rise to nontrivial dynamical effects. In particular, when the internal mode is excited, the interaction can trigger an instability in the transverse position of the string. Such an instability may have significant consequences, including the transfer of energy between distinct sectors of the effective theory.

The paper is organized as follows. In \Cref{Sec:AHCS} we introduce the Abelian Higgs model and describe the static vortex string solution. Next, in \Cref{Sec:general_discussion}
we outline the procedure for obtaining the low energy effective action for string excitations, including the zero energy fluctuations (Goldstone modes) and their interaction with the vortex bound state, which is discussed in \Cref{Sec:Seff+BS}. Details of the derivations are presented in \Cref{app:Derivations}. 
Finally, in \Cref{sec:numerics_effective}, we compare our semi-analytic results with those obtained from full field theory simulations. These are carried out using high-resolution lattice field theory techniques supplemented by Adaptive Mesh Refinement, enabling an accurate treatment of the multiple physical scales involved and ensuring reliable resolution of the distinct effects addressed in our numerical analysis. Details of the numerical techniques and the code can be found in \cite{Aurrekoetxea:2023fhl,Radia:2021smk} and the GRTL Collaboration website (\url{www.grtlcollaboration.org/}). We end up with a summary of our results and some concluding remarks in \Cref{Sec:Conclusions}.

We will set $c=1$ and use the mostly positive signature and a flat spacetime metric, ${\eta_{\mu \nu}={\rm diag}(-,+,+,+)}$.

\section{Abelian-Higgs cosmic strings}
\label{Sec:AHCS}
We will focus on the Abelian-Higgs model in 3+1 dimensions, with an action given by
\beq
\label{eq:FT-action}
S = - \int{\!\! d^{4}x \left[\frac 1 4 F_{\mu\nu}F^{\mu\nu}\!\!+\frac{1}{2} (D_{\mu} \phi)^*D^{\mu} \phi + \!\frac{\lambda}{8} \left(\phi^*\phi - \eta^2\right)^2\right]}~,
\eeq
where $F_{\mu\nu}=\partial_\mu A_\nu-\partial_\nu A_\mu$, is the field strength tensor and $D_\mu\phi=(\partial_\mu-ieA_\mu)\phi$ the gauge-covariant derivative of the charged scalar field. 
This model admits string-like solitonic solutions, the simplest one corresponding to an infinite straight static string first described by Nielsen and Olesen \cite{Nielsen:1973cs}. Such static solutions can be found by minimizing the static energy (per unit length) of the string. We will focus for simplicity on the critical case $(\lambda=e^2)$, in which the vortex strings become of Bogomolny-Prasad-Sommerfeld (BPS) type.

The solutions are topologically non-trivial, in the sense that the phase of the scalar field has a non-zero winding number around an axis corresponding to the zeros of the Higgs field $\phi$. Indeed, finiteness of the energy requires that $\phi\phi^*\to \eta^2$ and $D_i\phi\to 0$ when $\abs{x}\to \infty$ in the directions perpendicular to the string. 
If we choose the string axis to be aligned with the $z$-axis, then static solutions are given by the first order Bogomolny equations:
\begin{equation}
    D_x\phi\pm iD_y\phi=0,\qquad F_{xy}\pm \frac 1 2(\phi^*\phi-\eta^2)=0.
\label{BPS_eqs_init}
\end{equation}
It will be convenient to work with dimensionless quantities, which can be achieved by rescaling the fields accordingly,
\begin{equation}
    \phi\to\eta\phi,\quad A_\mu\to\eta A_\mu,\quad x^\mu\to\frac{1}{e\eta}x^\mu \,.
\end{equation}

Now let $\{\bar\phi(x,y),\bar A_\mu(x,y)\}$ be a nontrivial solution of the Bogomolny equations with unit topological charge. Then, a gauge can be chosen in which the vortex solution takes the form
\begin{align}
  \bar\phi(x,y)\equiv f(\rho)e^{i\theta},\quad   \bar A_\mu(x,y)\equiv\alpha(\rho)\partial_\mu\theta,
\end{align}
where $\rho^2=x^2+y^2$ and $\theta= \arctan(y/x)$ are the associated polar coordinates, with the profile functions $f$, $\alpha$ satisfying
\begin{equation}
    \rho f'=f(1-\alpha), \qquad \alpha'=\frac{1}{2}\rho (f^2-1),
\label{BPS_eqs}
\end{equation}
where primes denote derivatives with respect to $\rho$, and we supplement these equations with the standard boundary conditions ensuring regularity at the origin and finite energy, namely,
\begin{align}
    f(0)&=\alpha(0)=0,\notag\\
    \lim\limits_{\rho\to\infty}  f(\rho) &= \lim\limits_{\rho\to\infty}  \alpha(\rho) =1.
\label{BCns}
\end{align}
In Fig. \ref{fig:BPS_sols} (upper panel) we plot the solution of the system given by Eqns. (\ref{BPS_eqs}-\ref{BCns}).
As noted above, these solutions represent the lowest energy configurations within the relevant topological sector; here the vortex string of unit winding number. Many questions of physical interest, however, pertain to soliton dynamics away from these relaxed configurations. For example, transverse displacements of the string may be interpreted as excitations of the underlying field theory configuration that shift the vortex core. In addition, interactions with other strings and/or external fields can excite further internal modes of the soliton that may exist, leading to distortions of its profile. Such deformations can carry a non-negligible fraction of the total energy of the configuration and are therefore potentially important for a realistic description of string dynamics. 

We therefore begin by analyzing the spectrum of excitations of the solitonic string and the interactions among these modes, which constitute key inputs for constructing the effective theory governing string dynamics.

As for any stable topological defect whose formation breaks spatial translational invariance, the lightest excitations are the massless modes associated with rigid translations of the soliton in its transverse (co-dimension) directions. As we just mentioned these gapless modes manifest themselves as propagating wiggles along the string’s longitudinal directions. We will sometimes refer to these excitations as \textit{zero modes}.

A second class of excitations consists of bound states, i.e. massive modes localized in the soliton core. For the vortex string, the relevant bound state is commonly referred to as the {\it shape mode}, corresponding to a deformation of the transverse profile. The existence and number of such modes are model dependent; in the Abelian-Higgs theory, at least one shape mode is present \cite{Goodband:1995rt,Alonso-Izquierdo:2015tta}. The theory also admits scattering states, namely bulk excitations that interact with the soliton and provide channels for energy loss to the vacuum through radiation or other decay processes. In particular, massive string excitations decay gradually via nonlinear interactions into bulk scattering states \cite{Alonso-Izquierdo:2024tjc}. Although this decay is slow, the long-lived presence of these massive modes can influence the subsequent evolution of the string worldsheet and should therefore be taken into account.

\section{The effective action of a string}
\label{Sec:general_discussion}

Having described the underlying field theory soliton solution for our cosmic string model and the physical origin of its lowest energy excitations, we now turn to the construction of an effective action that captures the dynamics and interactions of these modes. We proceed systematically, organizing the analysis order by order in the excitation energy, first identifying the effective action for the massless modes and then incorporating higher energy modes.

\subsection{Effective action for the Goldstone mode revisited}

In order to derive the effective action for the string, we will make use of the well-known method of adapted coordinates (see Refs. \cite{Maeda:1987pd,Anderson:1997ip, Gregory:1988qv,Capovilla:1994bs,Anderson:book}). We start by considering an arbitrarily-shaped vortex string soliton, and let  ${x^\mu \equiv x^\mu(\sigma^A)}$ represent the points of the hypersurface swept out by the location of zeros of the Higgs field in spacetime. Such surface can be thought of as the worldsheet $\mathcal{W}$ of a string onto spacetime, parametrized by two worldsheet coordinates (one timelike and another spacelike), $\sigma^A$, $(A=0,1)$.  In four dimensions, $\mathcal{W}$ can be embedded as a (timelike) co-dimension 2 surface, hence at each point on the worldsheet we can choose two spacelike vectors, $n^\mu_i$, $(i=2,3)$, which are normal to the two tangent vectors $e^\mu_A\equiv \partial x^\mu/\partial \sigma^A$, i.e.
$
   \eta_{\mu\nu} n^\mu_i e^\nu_ A=0,
$
and orthonormal between them, 
$    \eta_{\mu\nu}n^\mu_i n^\nu_j=\delta_{ij}$.

With this choice, we can define a coordinate chart in which any given point in the neighborhood of the worldsheet can be parametrized as
\begin{equation}
   y^\mu(\xi)=x^\mu(\sigma^A)+u^i~n^\mu_i(\sigma^A)~,
\end{equation}
where $\xi^\mu=(\sigma^A,u^i)$.
Such a coordinate chart will be valid to describe points whose distance to the string is smaller than its curvature radius. These coordinates are also known as \emph{adapted} coordinates. A crucial remark is the fact that, although the choice of orthonormal basis $\{n_i^\mu,e^\mu_A\}$ is unique (up to local orthogonal rotations on the normal space), the same is not true for the coordinates $u^i$. In fact, there is still freedom for a reparametrization of the form $\tilde u^i(u ^j)$. We will choose the coordinate chart such that, for an arbitrarily shaped string configuration described by $\{\phi(x^\mu),A_\nu(x^\mu)\}$, the shape of the vortex is locally (in its own rest frame) given by the static solution, i.e. 
\begin{equation}
    \phi(x^\mu(\xi^\mu))\equiv \bar\phi(u),\quad A_\nu(x^\mu(\xi^\mu))\equiv \bar A_\nu(u),
    \label{zero_order_term}
\end{equation}
plus perhaps higher order corrections that come from the excitation of massive bound states, which we will not consider at this moment.

Our main goal is to reduce the four dimensional field theory action \eqref{eq:FT-action} to a two-dimensional effective action defined on the string worldsheet
\begin{equation}
      S^{(0)}_{\rm eff}=-\int_{\mathcal{W}}  d^2 \sigma\sqrt{-\gamma} \mathscr{L}_{\rm eff}.
\end{equation}

Where $\sqrt{-\gamma}$ is the determinant of the metric induced in the worldsheet, namely,
\begin{equation}
    \gamma_{AB}=g_{\mu\nu}e^\mu_A e^\nu_B,
\end{equation}
where $g_{\mu  \nu}$ denotes the $3+1$ dimensional spacetime metric. To derive the effective Lagrangian $\mathscr{L}_{\rm eff}$, we first note that the Lagrangian is a scalar density of rank one, so, when written in the new coordinates, we have
\begin{align}
    \lag[\phi(x),A_\mu(x)]=\sqrt{-g(\xi)}\mathcal{E}_\mathrm{BPS}(\rho)
\end{align}
 where $\mathcal{E}_\mathrm{BPS}(\rho)$ is the energy density of a BPS vortex, and we have explicitly remarked that it only depends on the normal coordinates through $\rho=\sqrt{u_i u^i}$.
 
We must then expand the metric determinant in the new coordinate set and integrate over transverse dimensions. Following \cite{Anderson:book,Anderson:1997ip} (see also \Cref{geo_review}), we can write:
\begin{equation}
    \sqrt{-g}=\sqrt{-\gamma}\Big[1+\bar{K}_1+\frac{1}{2}(\bar{K}_1^2-\bar{K}_2)\Big]
    \label{det_effmet}
\end{equation}
where $\bar{K}_n\equiv \Tr{(u^i\bm{K}_{i})^n}$, and $\tensor{K}{_{i}^B_A}$ is the extrinsic curvature tensor in the $i-$th direction. Furthermore, the axial symmetry of $\mathcal{E}_\mathrm{BPS}$ simplifies the transverse integrals

to arrive at 
\begin{equation}
    S^{(0)}_{\rm eff}=-\int d^2\sigma \sqrt{-\gamma}\Big[\mu +\nu\, \mathcal{R}\Big]
\label{Seff}
\end{equation}
where
\begin{equation}
    \mu=2\pi\int \mathcal{E}_\mathrm{BPS}(\rho) \rho~d\rho,\qquad \nu =\pi\int \mathcal{E}_\mathrm{BPS}(\rho) \rho^3 d\rho \,.
\end{equation}
Here $\mathcal{R}$ is the Ricci scalar of the worldsheet, and we have made use of the Gauss-Codazzi relation:
\begin{equation}
    \mathcal{R}=\delta^{ij}\Big[\tensor{K}{_{i}^A_A}\tensor{K}{_{j}^B_B}-\tensor{K}{_{i}^A_B}\tensor{K}{_{j}^B_A}\Big].
\end{equation}

The effective action \eqref{Seff} is not a new result. Indeed, it has been already derived multiple times in the literature, see e.g. \cite{Gregory:1988qv}, and is sometimes referred to as the `geometric part' of the effective action, as opposed to the `field theory' part that originates from further corrections to the field profile \cite{Anderson:1997ip}.

We will argue that this latter contribution is, in fact, not needed, as it effectively doubles the count of degrees of freedom. Instead, genuine profile deformations beyond the zero modes should be incorporated as additional massive excitations such as bound state modes localized on the string worldsheet which are harder to excite.

From this perspective, the effective action \eqref{Seff} is the \emph{complete} effective action that describes the dynamics of Goldstone degrees of freedom of a relativistic, field theory string defect, at least at the leading order.

This is consistent with the observation of \cite{Hinterbichler:2010xn} that no other DBI Galileon terms exist for even codimension $N\geq 2$ in $d=4$\footnote{In the particular case of codimension 2, another term may appear \cite{Hinterbichler:2010xn}, but it is not independent of the Ricci scalar term if the background is maximally symmetric \cite{Garoffolo:2025igz}. We do not get such term as we are always considering flat spacetime.}. In particular, we should not observe any deviations from the dynamics predicted by the Nambu-Goto equations of motion coming from higher curvature terms, as the integral of $\mathcal{R}$ over the two-dimensional string worldsheet is a topological invariant, so that this term does not affect the dynamics derived from \eqref{Seff}.

\subsection{Effective action including a bound state}
\label{Sec:Seff+BS}

\begin{figure}[t]
    \centering
    \includegraphics[width=\linewidth]{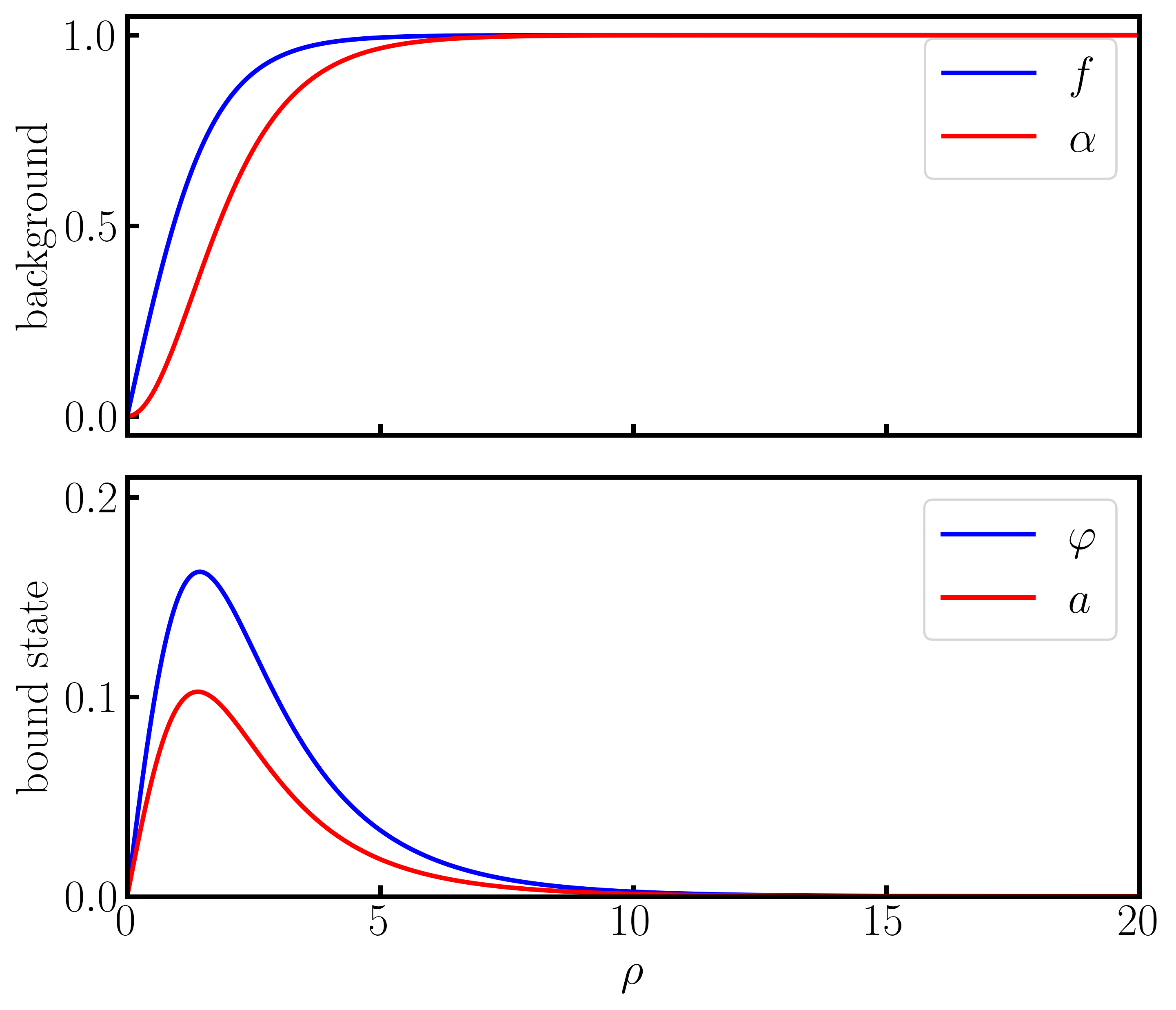}
    \caption{Top: Scalar and vector radial profile for the critical vortex. Bottom: Scalar and vector part of the first bound state solution of the linearized perturbation equations for the critical vortex. The modes are normalized using the standard $L^2$ norm on the plane.}
    \label{fig:BPS_sols}
\end{figure}

Let us now consider, in addition to the massless Goldstone degrees of freedom, further excitations of the string. One of the simplest possible choices consists of the excitation of the lowest energy, linearly stable bound state of a straight string (or equivalently, the bound state of a $2+1$ Abelian-Higgs vortex). Indeed, such bound state preserves the axial symmetry of the static configuration, and its spatial profile can be found by solving the equation of motion for linearized perturbations on the vortex background with the ansatz \cite{Alonso-Izquierdo:2024tjc}:
\begin{align}
            \phi(x,t)&=[f(\rho)+\epsilon ~\varphi(\rho)\cos(\omega t)]e^{i\theta},\notag\\ A_\mu(x,t)&= [\alpha(\rho)+\epsilon ~\rho~a(\rho)\cos(\omega t)]\partial_\mu\theta,
\end{align}
which, at first order in $\epsilon$, yields the following eigenvalue problem:
\begin{equation}
    \mathcal{L}\mqty(\varphi_n\\a_n)=\omega_n^2\mqty(\varphi_n\\a_n),
\end{equation}
where $\mathcal{L}$ is the second order differential operator given by \cite{Alonso-Izquierdo:2024tjc}:

\begin{equation}
  \mathcal{L}=-\qty(\dv[2]{}{\rho}+\frac{1}{\rho}\dv{}{\rho})\bm{1}+\mqty(U(\rho)&W(\rho)\\[2mm]W(\rho)&V(\rho)),  
\end{equation}
where $\bm{1}$ is the $2\times2$ identity matrix, and
\begin{align}
    U(\rho)&=\frac{3 f(\rho)^2}{2}+\frac{(\alpha(\rho)-1)^2}{\rho^2}-\frac{1}{2},\\
    V(\rho)&=f(\rho)^2+\frac{1}{\rho^2},\quad
    W(\rho)=-2\frac{f(\rho)(\alpha(\rho) -1)}{\rho}.
\end{align}
In the critical case, it turns out that there is a single bound state solution of the above eigenvalue problem, with $\omega_1^2\equiv\omega^2 \approx 0.77$ \cite{Goodband:1995rt,Alonso-Izquierdo:2015tta}, and whose respective scalar and vector radial profiles are plotted in Fig. \ref{fig:BPS_sols}.

Since such mode only modifies the shape of the vortex, but keeps intact the residual gauge and Poincaré symmetries, we will refer to it simply as the `shape mode' of the vortex, or equivalently, of the string.

Therefore, we will be interested in modifying \eqref{zero_order_term} with an additional scalar field living on the string worldsheet, $\chi(\sigma^A)$, that parametrizes the amplitude of the excited shape mode :
\begin{align}
    \phi(x^\mu(\xi^\mu))&= \bar{\phi}(u)+\chi(\sigma^A)\varphi(u),\notag\\
    A_\nu(x^\mu(\xi^\mu))&= \bar{A}_\nu(u)+\chi(\sigma^A)a_\nu(u).
    \label{ansatz_perturb}
\end{align}

Then, the field theory action can be expanded in powers of this scalar field:
\begin{equation}
    S=S^{(0)}_{\rm eff}+\int \left(\delta^{(1)}S\right)\chi~ d^4x+\frac{1}{2}\int \chi \left(\delta^{(2)}S\right) \chi~d^4x+\order{\chi^3}
    \label{S_expansion}
\end{equation}
where
\begin{align}
    \delta^{(1)}S&=\varphi\frac{\delta S}{\delta\phi}+a_\mu\frac{\delta S}{\delta A_\mu},
\label{dS1}\\
      \delta^{(2)}S=\varphi^2\frac{\delta^2 S}{\delta\phi^2}&+a_\mu a_\nu\frac{\delta ^2S}{\delta A_\mu\delta A_\nu}  + 2a_\mu \varphi\frac{\delta ^2S}{\delta A_\mu\delta \phi},
\end{align}
and functional derivatives must be evaluated on the zeroth-order solutions, $(\bar\phi,\bar A_\mu)$.

The first term in the rhs of Eqn. \eqref{dS1} is proportional to the equations of motion for the scalar field, and the same is true for the second term but for the gauge field.  In \Cref{app:Derivations} we show that, when written in adapted coordinates and evaluated on the BPS solutions, these terms can be written as,
\begin{equation}
    \delta^{(1)}S=\Tr{\frac{\bm{K}^i}{\bm{1}+u_l\bm{K}^l}}\qty(\nabla_ i\bar{\phi}\varphi+\eta^{jk}\bar{F}_{ij}a_k)~,
\label{dS_BPS}
\end{equation}
which is the higher codimension generalization of the tadpole term that appears in the case of domain walls \cite{Blanco-Pillado:2024bev,Zia:1985gt}.

We may now expand \eqref{dS_BPS} in powers of the transverse coordinates $u^i$. As it turns out (see \Cref{app:Derivations}) the expansion truncates, generating just the following term in the effective action after integrating along transverse directions:
\begin{equation}
    S^{(1)}_{\rm eff}=\int \left(\delta^{(1)}S\right) ~\chi ~d^4x=-\kappa \int \sqrt{-\gamma}\mathcal{R}(\sigma)\chi(\sigma)d^2\sigma~,
    \label{coupling_R}
\end{equation}
where $\kappa$ is a coupling constant given by
\begin{equation}
\kappa=\pi\int_0^\infty\qty(\rho^2 f'\varphi+\rho\,\alpha'a)d\rho\approx -0.7646\pi~.
\end{equation}
At second order in the field amplitude, the effective action is that of a free massive scalar with a curvature dependent mass, 
\begin{equation}
    S^{(2)}_{\rm eff}\approx -\frac{1}{2}\int \!\!\sqrt{-\gamma}~\chi~\qty[-\gamma^{AB}\nabla_{AB}+\omega^2(1+c_2\mathcal{R})]~\chi~d^2\sigma+\cdots
    \label{dS2_eff}
\end{equation}
where the extra terms are suppressed by curvature, as shown in \Cref{app:Derivations}, and $c_2$ is given by \eqref{c2}.

Putting all terms together we arrive at the following effective action
\begin{equation}
    S_{\rm eff}\!\approx \!-\!\!\int \!\!\!\sqrt{-\gamma} \!\left[ \mu \!+ \!\frac{1}{2} (\partial^A \chi)^2 \! +\!\frac{ \omega^2}{2} (1\!+\!c_2\mathcal{R})  \chi^2\! +\! \kappa  \chi\mathcal{R} \right]\!d^2\sigma + \cdots
\label{EFT=NG+scalar}
\end{equation}
where the dots include 
all other higher order terms that we have neglected on the expansion of $S^{(2)}_{\rm eff}$ and $S^{(n)}_{\rm eff}$ for $n=  3,4$.

In summary, the procedure described above provides a direct derivation, from the underlying field theory solution, of an effective action for the Abelian-Higgs string that incorporates the Goldstone sector together with the leading massive excitation. The resulting theory consists of Nambu-Goto dynamics for the string, supplemented by a massive scalar field propagating on the worldsheet. In addition, our construction reveals a non-minimal interaction between the worldsheet geometry and the scalar mode. At the lowest order this manifests itself as a linear coupling of the scalar to the intrinsic Ricci scalar built from the induced metric. As we demonstrate in the following sections, this term gives rise to distinctive features in the dynamics of the excited string.

\section{Comparison with field theory simulations}
\label{sec:numerics_effective}

In order to test the derived effective action, we compare the dynamics of strings predicted by the action given by Eqn. (\ref{EFT=NG+scalar}) and that obtained from a direct numerical field theory simulation. In order to solve for the field theory dynamics, we implement the equations of motion obtained from Eqn. \eqref{eq:FT-action}  in \texttt{GRDzhadzha} \cite{Aurrekoetxea:2023fhl}, as described in \cite{Zilhao:2015tya,Helfer:2018qgv}. This is a code by the GRTL Collaboration that incorporates Adaptive Mesh Refinement (AMR) to accurately resolve the different scales in the problem: the string core and the length of the string.

To enable a meaningful comparison between these two descriptions of string dynamics, the field theory initial data should closely match the assumptions implicit in the worldsheet effective theory. This requirement restricts the class of initial conditions that can be implemented straightforwardly in our field theory simulations. Nevertheless, a large, mildly curved circular loop provides a natural analogue of an initially static circular string in the effective action framework. In what follows, we adopt this configuration as our first benchmark for testing the worldsheet effective theory.

\subsection{Collapsing circular string loops}

\begin{figure}[t]
\includegraphics[width=\linewidth]{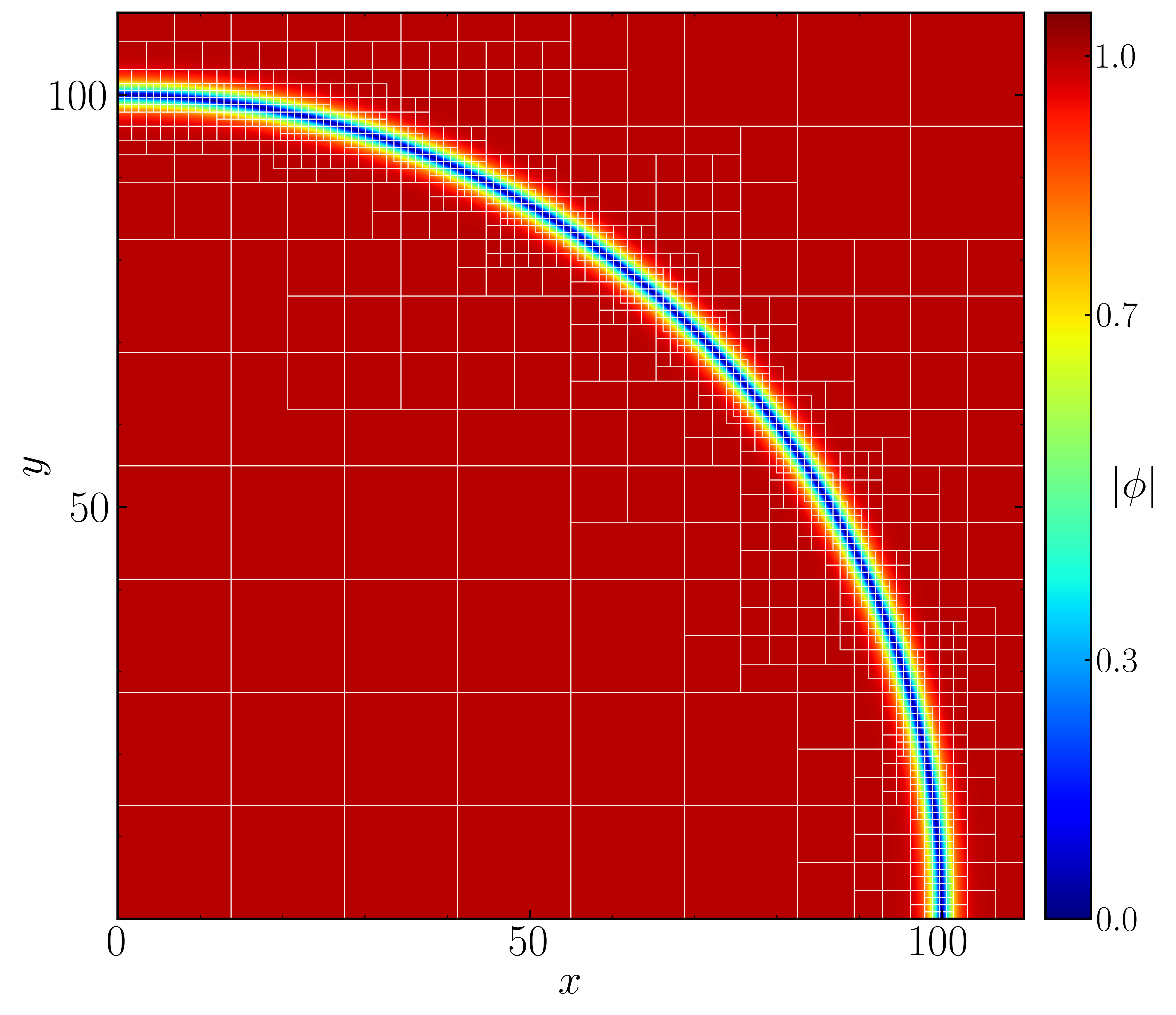}
    \caption{Initial data for a loop with $R_0=100$. The white lines depict the 6 levels of adaptive mesh refinement, with each box containing $32^3$ number of grid points. We use reflective boundary conditions in $x$, $y$ and $z$.}
    \label{fig:loop_2d}
\end{figure}

In this section, we take as initial data a circular string loop whose local transverse (radial) profile is identified with that of the relaxed, lowest energy straight string solution. For loops with radius much larger than the string thickness, this constitutes an excellent approximation to a string initially at rest. This setup is the simplest nontrivial configuration with nonzero curvature and its subsequent evolution should lead to a collapsing circular loop. The presence of large Lorentz contraction in this evolution indicates the need of a fine grid in order to accurately simulate this process. This is where the AMR techniques we use here prove to be important.

Such collapses in the Abelian-Higgs model have been simulated previously, both including gravitational backreaction \cite{Helfer:2018qgv,Aurrekoetxea:2020tuw,Aurrekoetxea:2023vtp} (with AMR) and in the purely field theory case (without AMR) in \cite{Nagasawa:1994md}\footnote{See also Refs. \cite{Drew:2019mzc,Drew:2022iqz,Drew:2023ptp} for flat-space simulations of global strings with AMR.}. However, the impact of curvature induced corrections to Nambu-Goto (NG) dynamics has not been systematically explored. Addressing this issue is a central goal of the present work. Here we closely follow the setup of Refs.~\cite{Helfer:2018qgv,Aurrekoetxea:2020tuw}, but neglect gravitational backreaction (i.e., we work in a fixed, non-dynamical flat spacetime background).

Figure~\ref{fig:loop_2d} illustrates a representative initial scalar field configuration for a circular loop initially at rest. The figure also displays several levels of the adaptive mesh refinement, highlighting the need for high resolution in the vicinity of the string core and progressively fewer refinement levels farther away. This requirement becomes even more stringent during the subsequent evolution: as the loop contracts and attains high velocities, accurately resolving the dynamics demands substantially increased refinement. Given the symmetry of the system, we use reflective boundary conditions along all $x, y, z$, so that we only evolve one octant of the full box. The radius of this loop is $R_0=100$, in a box $L=165$ and $N=96^3$ number of cells in the coarsest grid and $6$ additional levels of refinement. The finest resolution is then $dx\approx 0.025$.

\subsubsection{Lowest order comparison. The Nambu-Goto case}

One of the main conclusions that follows from the effective action derived in the previous section is that, at leading order, in the absence of excitations of the bound mode, the string dynamics are governed solely by the Nambu-Goto action, with no additional higher-derivative corrections. To test this prediction, we compare the collapse of an initially static loop in the full field theory with the corresponding evolution obtained from the Nambu-Goto description. To this end, we parametrize the loop worldsheet in Cartesian coordinates as follows:
\begin{equation}
    X^\mu(\sigma,\tau)=(\tau,R(\tau)\cos(\sigma), R(\tau)\sin(\sigma),0),
\end{equation}
so the NG equations reduce to a single equation for the loop radius 
whose solution for a static initial condition is given by \cite{Vilenkin:1981kz}
\begin{equation}
    R(\tau)=R_0\cos\left(\frac{\tau}{R_0}\right)
    \label{eq:NG}
\end{equation}
with $\tau\in\left[0,\, \pi R_0/2\right]$.

\begin{figure}[t]
\includegraphics[width=\linewidth]{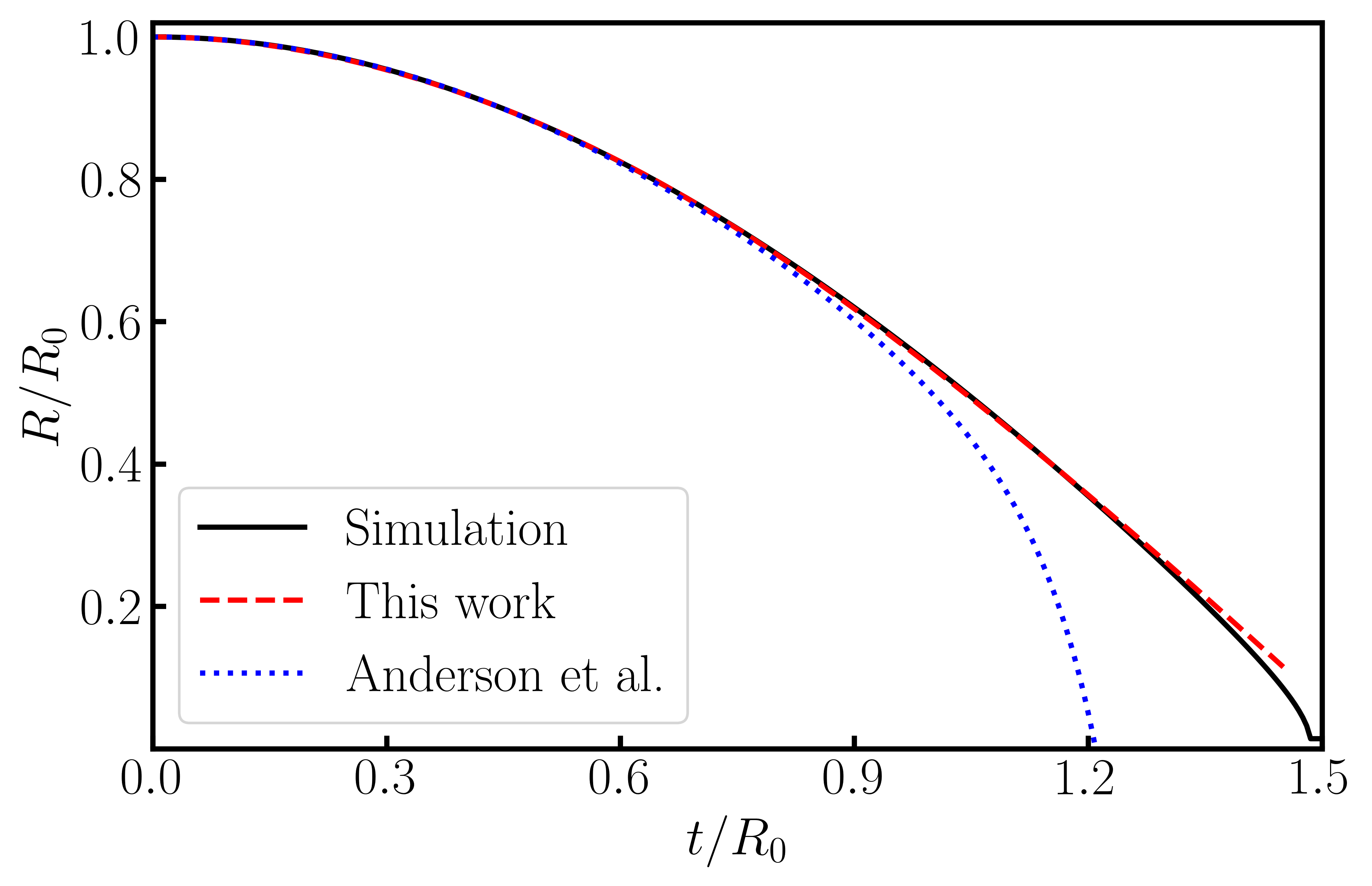}
    \caption{Comparison of the evolution of the radius of the circular string loop as a function of time, with $R_0=10$. The black solid line corresponds to numerical solution. The red dashed line corresponds to the NG solution, while the dotted blue includes the correction in Eqn. \eqref{eq:Anderson}.}
    \label{fig:loop_rad_10}
\end{figure}

\begin{figure*}[t]
\includegraphics[width=\linewidth]{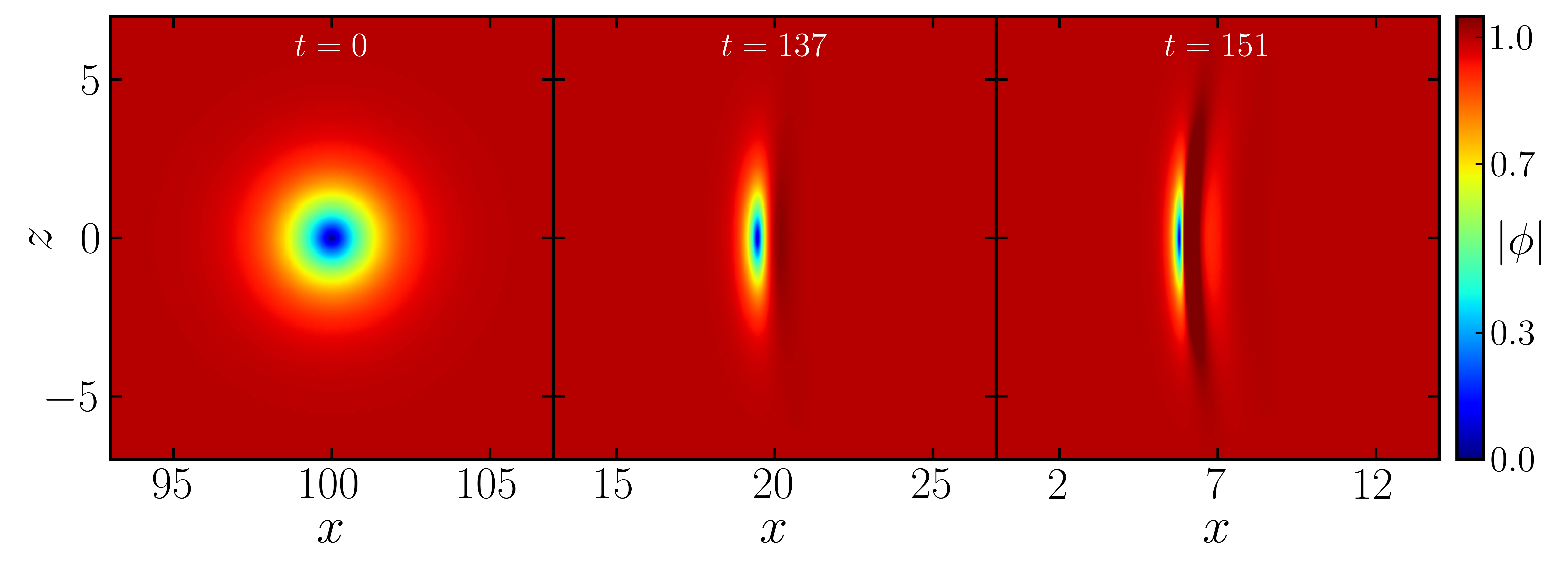}
    \caption{Slice across $y=0$ showing the $x-z$ plane evolution as the loop with initial radius $R_0=100$ collapses.}
    \label{fig:profiles}
\end{figure*}

We claim that the above result is valid to describe the position of the string center for a collapsing loop right up to the collapse time, with no further corrections coming from higher curvature terms in the effective action. This is in contrast with previous theoretical results on the dynamics of string defects. In particular, it was claimed in \cite{Anderson:1997ip} that due to higher curvature terms in the effective action, the NG solution \eqref{eq:NG} gets modified as $R(\tau)+\delta R(\tau)$, with
\begin{align}
\delta R(\tau) &= 32\,\epsilon^{4}\zeta \Bigg(
\frac{7}{40}\sec^{5}\!(\tau) + \frac{1}{60}\sec^{3}\!(\tau) \notag\\
&\qquad + \frac{1}{15}\sec (\tau) - \frac{31}{120}\cos(\tau)
- \frac{\tau}{8}\sin(\tau)
\Bigg), \label{eq:Anderson}
\end{align}
where $\epsilon$ is the inverse initial loop radius in units of the string width $\delta_s=\pi\mu^{-1}$, and $\zeta$ is a numerical constant of order one in units of $\delta_s$, which we take to be $\zeta=-1.1\pi$, based on Fig. 7 in Ref. \cite{Anderson:1997ip}.

To elucidate this discrepancy, we simulate the collapse of a loop with initial radius $R_0=10$. We employ a computational domain of size $L=16$ with $N=128^3$ cells on the coarsest grid. We further include three levels of adaptive refinement, achieving a finest spatial resolution of $dx=0.015$ (To be compared to the thickness of the string in this units which is of the order of $\delta_s\approx 1$). From the resulting data, we extract the string position by locating the core, defined as the zero of the Higgs field, and determining its center via interpolation on the grid.

Figure~\ref{fig:loop_rad_10} compares the simulated time evolution of the loop radius with the prediction of the Nambu-Goto action (Eqn.~\eqref{eq:NG}) and with the corrected expression proposed in Ref.~\cite{Anderson:1997ip} (Eqn.~\eqref{eq:Anderson}). We find no significant deviation from the NG prediction until the late stages of the collapse, in agreement with the expectations from our effective action analysis.

It is instructive to examine the field configurations during the collapse. Part of the motivation underlying Eqn.~\eqref{eq:Anderson} is that nonzero worldsheet curvature should induce deformations of the local field profiles. This expectation is borne out in Fig.~\ref{fig:profiles}, where one observes deviations that cannot be accounted for by a simple Lorentz boost of the static solution. Nevertheless, these deformations do not translate into an observable modification of the trajectory of the string core as one can see in Fig.~\ref{fig:loop_rad_10}.

In our framework, this is understood as a coordinate effect: the apparent profile distortions arise from the transformation between the coordinates of an asymptotic (laboratory frame) observer and the adapted coordinates used to construct the effective action. An analogous phenomenon was discussed in detail for collapsing domain walls, where a direct comparison between the two coordinate descriptions was carried out \cite{Blanco-Pillado:2024bev}. 

\subsubsection{Including the massive worldsheet scalar}

We now discuss the potential impact of a massive scalar degree of freedom on the worldsheet 
of the collapsing loop. At first sight, one might expect no effect, since our initial data are constructed to approximate a maximally relaxed string profile, so that the initial amplitude of the scalar field associated with the shape mode deformation is negligible. However, this expectation is not strictly correct according to our effective action. Looking at Eqn. (\ref{EFT=NG+scalar}) we see that the scalar couples non-minimally to the worldsheet Ricci scalar, which induces an effective potential even when the field amplitude vanishes. Detecting a nontrivial signal in this regime would therefore provide a sharp test of the effective description.

Extracting the scalar amplitude directly from field theory simulations is, however, technically challenging. A reliable projection onto this mode is already nontrivial for straight vortices, and becomes substantially more difficult for a curved string that accelerates to relativistic velocities. We therefore focus on an observable that can be robustly obtained from the simulations: the string position. Specifically, we search for the backreaction on the position of the core due to the massive mode present in the collapsing loop.

\begin{figure}[t]
\includegraphics[width=\linewidth]{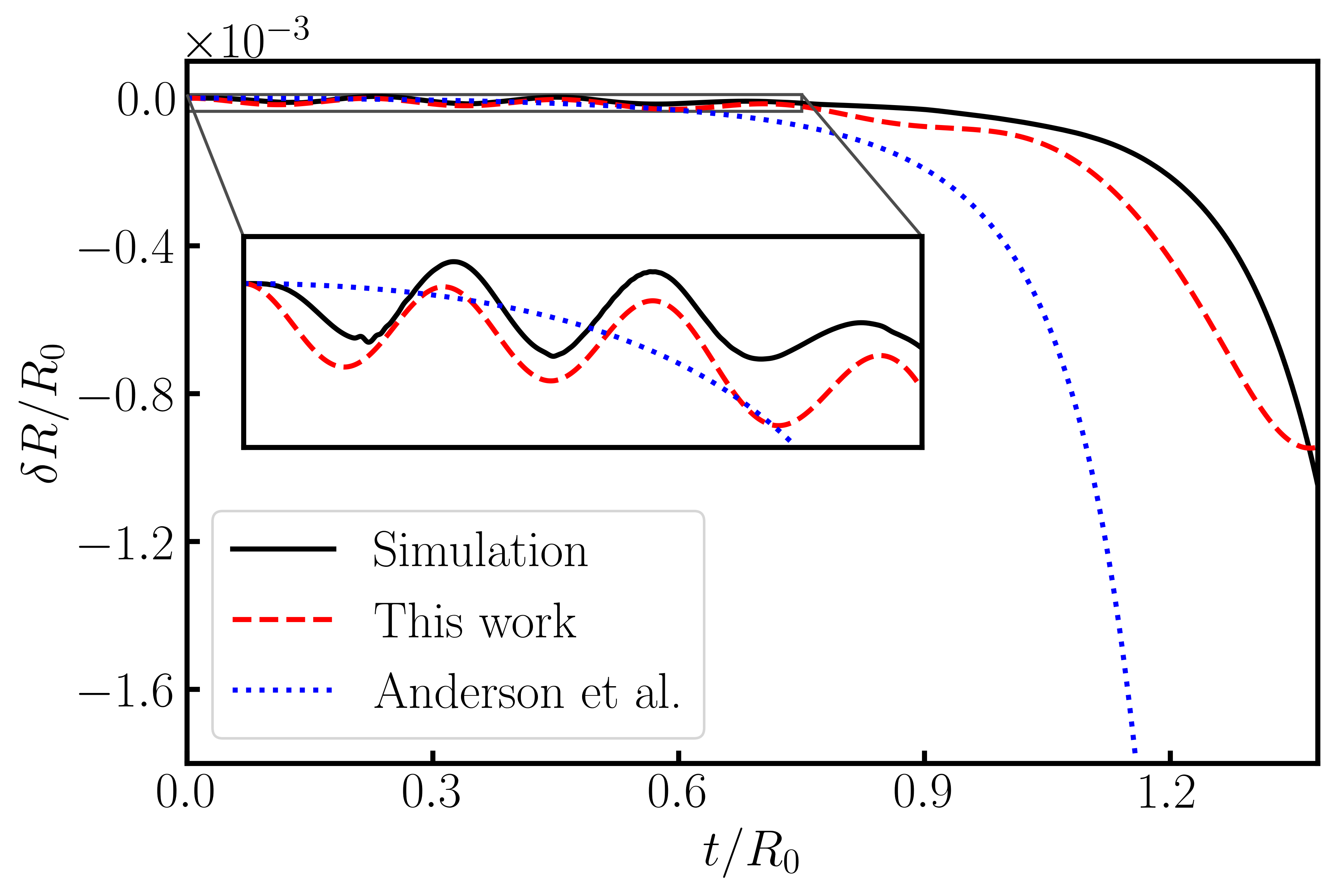}
    \caption{Deviation from the Nambu-Goto solution for a loop with initial radius $R_0=32$: predictions from our effective model compared with those of \cite{Anderson:1997ip} and with the field theory simulation.}
    \label{fig:loop_rad_32}
\end{figure}

To this end, we plot in Fig.~\ref{fig:loop_rad_32} the deviation of the loop radius from the strict Nambu-Goto evolution given by Eqn.~(\ref{eq:NG}). The field theory simulation shown there corresponds to $R_0=32$, performed in a domain of size $L=64$ with $N=32^3$ cells on the coarsest grid and five refinement levels, yielding a finest resolution of $dx = 0.0625$. The deviation remains extremely small, confirming that the Nambu-Goto solution provides an excellent approximation. Nonetheless, we observe small oscillations superimposed on this deviation, which qualitatively agree well with the expectations from the effective action also shown in the figure\footnote{We choose an intermediate initial radius, $R_0=32$, to balance competing requirements: it must be large enough to allow the collapse to last several scalar field oscillation periods, yet small enough that the deviation amplitude remains measurable.}. This provides strong evidence that the derived worldsheet theory captures the coupling between the Goldstone sector and the leading massive excitation.

What is the regime of validity of the effective model? As argued in \Cref{app:Derivations}, the effective action for the bound state amplitude Eqn. \eqref{dS2_eff} contains just the first nontrivial terms of an expansion in powers of worldsheet curvature invariants, and therefore we should expect that it breaks down whenever the \emph{worldsheet} curvature radius becomes comparable to the typical size of the string. To estimate this point, we may define the typical curvature radius of the worldsheet as $1/\sqrt{\mathcal{R}}$, 
and therefore we expect the effective action including the bound state contribution to break down around the instant when $1/\sqrt{\mathcal{R}}\simeq 1$.
This time corresponds to the point at which the effective theory prediction and the simulation results begin to exhibit a pronounced discrepancy in Fig.~\ref{fig:loop_rad_32}, at $t/R_0\approx 1.4$.

Having found strong evidence that worldsheet curvature induces a nontrivial excitation of the bound scalar mode even when its initial amplitude is negligible, we can probe this coupling more directly by initializing the field theory simulation in a controlled excited state. To this end, we construct initial data for a curved string with a prescribed perturbation corresponding to a shape mode amplitude $\chi_0=0.1$. Such initial excitation produces a measurable backreaction on the string trajectory, which can be compared quantitatively with the predictions of the effective action.

\begin{figure}[t]
\includegraphics[width=0.97\linewidth]{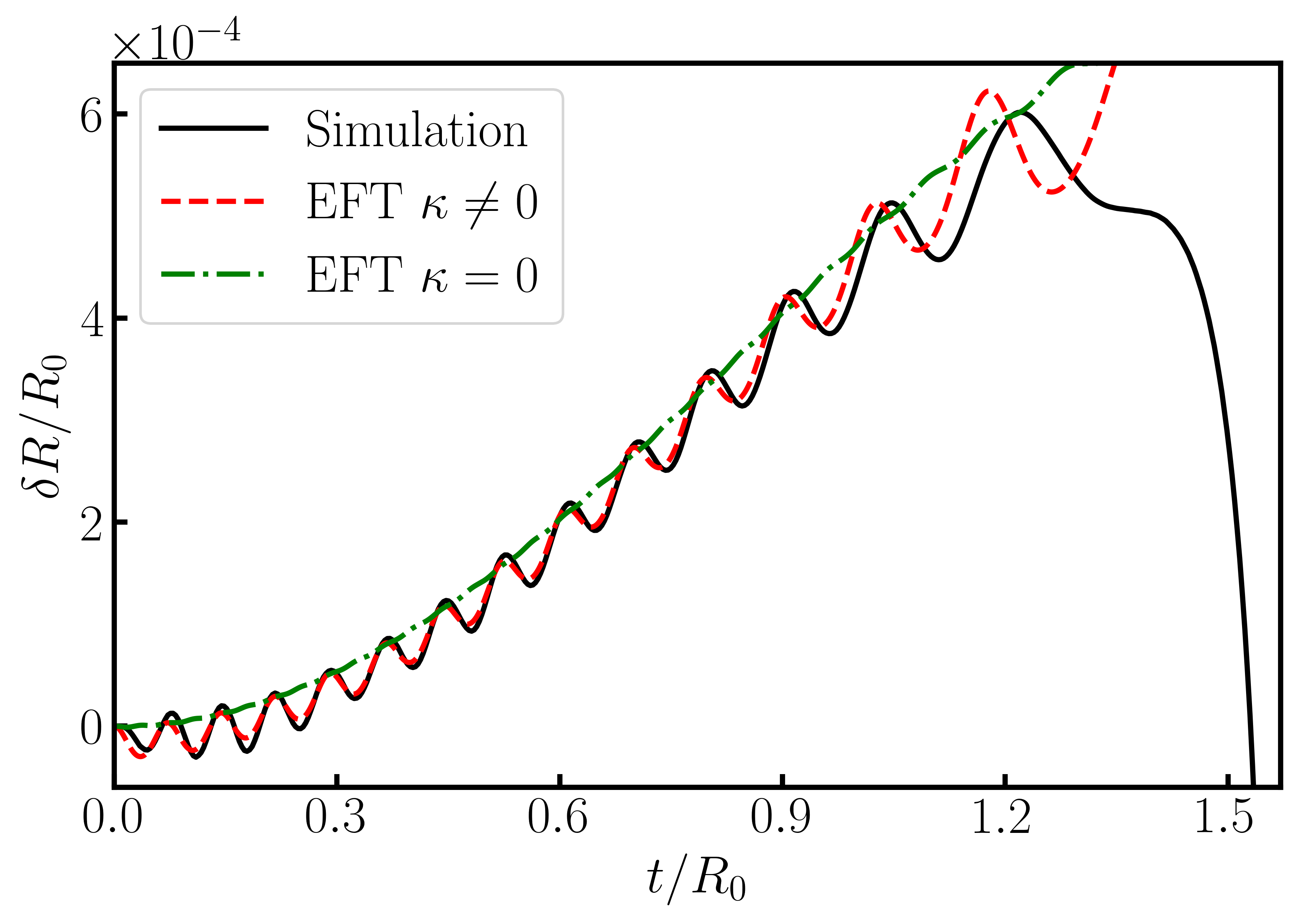}
    \caption{Deviation from the Nambu-Goto solution as predicted by our effective model. The initial radius of the loop is $R_0=100$, with an initial amplitude of the shape mode $\chi_0=0.1$.}
    \label{fig:loop_rad_100}
\end{figure}

Figure~\ref{fig:loop_rad_100} shows the deviation of the string core position from the Nambu-Goto collapse, extracted from a simulation with $R_0=100$, $L=165$, $N=96^3$ and $6$ levels of refinement, yielding a finest resolution $dx = 0.025$. We compare these results with the effective theory prediction. To disentangle the backreaction of a minimally coupled free massive worldsheet scalar from the effects of the non-minimal coupling, we display both scenarios. The comparison demonstrates that the curvature-induced coupling is required to reproduce the field theory evolution. It also provides a nontrivial validation of the effective action coefficients since small variations of these parameters would lead to appreciable changes in the predicted trajectory.

Taken together, these numerical tests support the presence of the non-minimal coupling of the bound mode in the effective action. In the next section we investigate one of its most striking consequences: a parametric instability triggered by excitations of the bound state on an otherwise straight string.

\subsection{Parametric instability of the excited local string}

We now consider a different class of initial data and study an infinitely long, straight string oriented along the $z$ direction. We introduce a small but finite, spatially uniform excitation of the core-bound mode (the shape mode), described by an amplitude $\chi(t)$. Given this configuration, we will investigate the subsequent evolution of the string position. As we show below, the dynamics generically lead to an amplification of a particular transverse perturbation of the string core, corresponding to a transfer of energy from the bound mode (shape mode) to the Goldstone (translational, zero mode) sector in a process described by a parametric instability. 

In fact, we will now demonstrate that this effect is naturally captured by the effective theory at the level of the linearized equations of motion for transverse wiggles. To this end, we consider a small deformation along the string that is confined to a fixed transverse plane (for definiteness, the $y - z$ plane i.e., $x=0$) and takes the form of a standing wave with wavelength $\lambda_0=2\pi \omega_0^{-1}$. The corresponding worldsheet embedding can be parametrized as follows:
\begin{equation}
    X^\mu=(t,0,\psi(t)\cos(\omega_0 z),z).
    \label{zero_pert}
\end{equation}

In the small-amplitude limit, the effect of the shape mode, $\chi(t)$ is to generate an approximately periodic modulation of the effective restoring force for $\psi(t)$. As a result, the early time evolution of $\psi(t)$ is well described by a Mathieu equation of the form:
\begin{equation}
    \ddot{\psi}(t)=-\omega_0^2\qty[1-\frac{2\kappa m_s}{\mu}\chi_0\cos(\omega t)]\psi(t)
\label{Mathieus}
\end{equation}
where we have used the harmonic approximation for the dynamics of the shape mode amplitude, 
\begin{equation}
    \ddot{\chi}+\omega^2\chi=0\implies \chi(t)=\chi_0\cos(\omega t),\quad m_s=\omega^2.
\end{equation}
Mathieu's equation \eqref{Mathieus} presents an instability whenever the following condition is satisfied  \cite{Kovacic2018MathieusEA}: \begin{equation}
    \frac{\omega_0}{\omega}=\frac{k}{2}, \quad k\in \mathbb{N}
\end{equation}
which implies the existence of a parametric instability in the coupled dynamics of the shape and zero modes of the string.

\begin{figure*}[t]
    \centering
\href{https://youtu.be/0qi1brTlTSg}{\includegraphics[width=\textwidth]{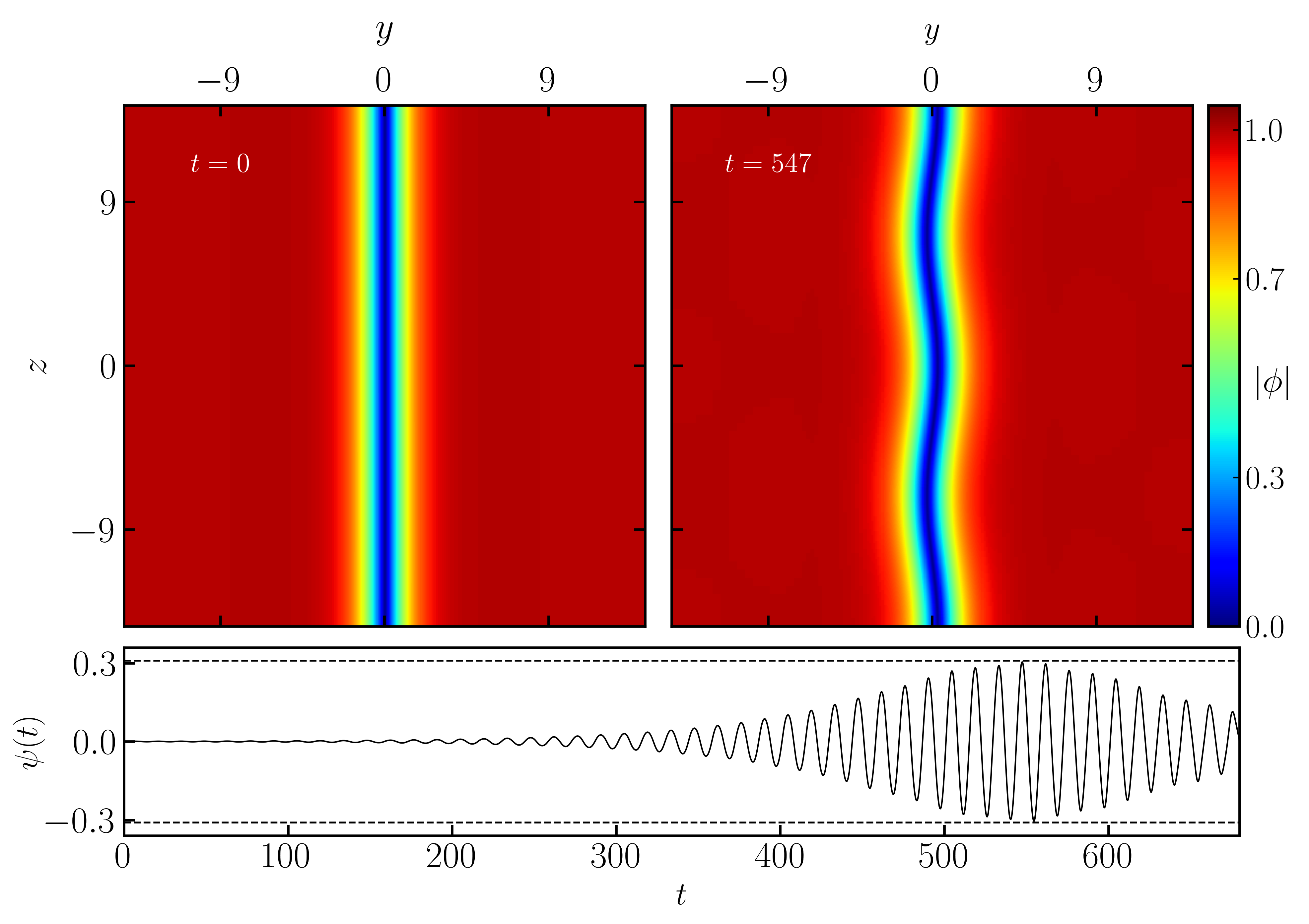}%
    }
    \caption{String profile in the $x=0$ plane. At $t=0$ (left panel) the shape mode is excited with initial amplitude $\chi_0=0.2$ and the zero-mode with initial amplitude $\psi_0=10^{-3}$. At a later time ($t\approx 550$) the zero mode is clearly excited due to the instability discussed in the text. Bottom panel: Position of the string center at $z=0$ as a function of time. This is equivalent to the amplitude of the zero mode. The horizontal dashed line shows the analytical prediction for the maximum zero-mode amplitude from energy conservation; see the main text. A movie of the simulation can be found in \url{https://youtu.be/0qi1brTlTSg}.}
    \label{fig:loop_2d_panel}
\end{figure*}
In order to reproduce such parametric instability in the field theory simulation, we must initialize the fields in a configuration with both the shape mode and the zero mode excited, the latter with an amplitude that is a function of the spacelike coordinate in the string worldsheet, $\Gamma^i$, according to Eqn. \eqref{zero_pert},
\begin{align}
    \phi=& \bar\phi+\chi\varphi+\Gamma^i\delta_i\bar\phi+\Gamma^i\chi\delta_ i\varphi,\notag\\
    A_\nu=& \bar A_\nu+\chi a_\nu+\Gamma^i\delta_i\bar A_\nu+\Gamma^i\chi\delta_ ia_\nu.
    \label{full_pert}
\end{align}
where $\delta_i$ denotes generically the linearized action of the translational zero mode in the $i$-th direction on the corresponding field. We remark that such action is not just a naïve translation as in the global string case, but the result must satisfy the Gauss law constraint, 
which implies \cite{Tong:2013iqa,Miguelez:2025}:
\begin{align}
    \delta_ i\phi=& D_i\phi\equiv \partial_i\phi-i\bar A_i\phi,,\notag\\
    \delta_ iA_\nu=& F_{i\nu}\equiv \partial_ i A_\nu-\partial_\nu A_i .
    \label{zero_linear}
\end{align}
Let us now consider the simple case in which the string zero mode is excited with an amplitude given by $\Gamma^i(t=0,z)=(0,\psi_0\cos(\omega_0 z))$, and the shape mode is uniformly excited, $\chi(t=0,z)=\chi_0$, such that the complete expression satisfies the gauge constraint to linear order in the perturbation amplitudes:
\begin{align}
    \phi=& \eta (f+\chi_0\varphi)e^{i\theta}+\psi_0\cos(\omega_0 z) D_z\bar\phi,\notag\\
    A_j=& [\alpha+\rho\chi_0 a]\partial_j\theta+\psi_0\cos(\omega_0 z)\bar F_{zj}.
    \label{part_pert}
\end{align}

where
\begin{align}
    D_z \bar\phi
    =\Big[\sin\theta\partial_\rho -i\frac{\cos\theta}{\rho}(\alpha-1)\Big]fe^{i\theta},
\end{align}

and 
\begin{equation}
    \bar F_{ij}=\epsilon_{ij}\frac{\alpha'}{\rho}.
\end{equation}

We perform a field theory simulation with $L=28.6$, $N=64^3$ and three levels of refinement, so that $dx\approx 0.05$. The initial amplitudes of the shape and zero modes are given by $\chi_0\equiv\chi(t=0)=0.2$ and $\psi_0\equiv\psi(t=0)=10^{-3}$. The top panel of Fig.~\ref{fig:loop_2d_panel} presents two representative snapshots of the string profile at different stages of the evolution. The first shows the nearly straight string at $t=0$, with a spatially uniform excitation of the bound (shape) mode. The second, taken at a later time, clearly exhibits the growth of the transverse deformation. The lower panel displays the extracted position of the string core along the line at the intersection of the $x=0$ and $z=0$ plane, illustrating the initial amplification of the transverse displacement as a function of time.

As we see in the figure, this growth is not indefinite and is eventually saturated once additional effects become important. From the field theory perspective, the amplification must cease because the energy initially stored in the shape mode is finite: once it has been transferred to the transverse mode, further growth is energetically forbidden. The effective theory description reaches the same conclusion: as the transverse amplitude increases, non-linear terms, neglected in the linearized analysis leading to the Mathieu equation, become relevant and invalidate the simple parametric instability approximation, thereby cutting off the exponential growth.

The maximum amplitude of the zero mode can be straightforwardly estimated using the previously mentioned argument based on energy conservation. At quadratic order, the energy per unit length stored in the shape mode and in the zero mode is, respectively: $E_s=  \omega^2 \chi^2/2$ and $E_z=\pi  \omega_0^2 \psi^2/4$ . Assuming that the energy of the shape mode is transferred to the zero mode due to parametric instability, we finally obtain: $\psi_\mathrm{max}=2 \sqrt{2/\pi}\chi_0$. For an initial shape mode amplitude $\chi_0=0.2$, we obtain a maximum zero-mode amplitude $\psi_\mathrm{max}=0.31$. As shown in the lower panel of Fig.~\ref{fig:loop_2d_panel}, this prediction is consistent with the field theory simulations.

Finally, we would like to emphasize that the nonzero initial amplitude assigned to the zero mode (i.e., the transverse displacement) is introduced solely for practical convenience, as it facilitates the extraction of the relevant observables from the simulations. The underlying mechanism does not rely on a finely tuned initial deformation: the analysis implies that any sufficiently small perturbation whether originating from numerical roundoff, discretization effects, or generic initial fluctuations will seed the instability, typically in an arbitrary transverse direction.

\section{Conclusions}
\label{Sec:Conclusions}
In this paper we have revisited the effective action for the low energy dynamics of a local string in the Abelian-Higgs model. We have provided both analytical and numerical evidence of the absence of higher order terms in the tree level effective action for the Goldstone modes apart from the NG term, contrary to previous claims in the literature.

The absence of higher curvature corrections in the effective action of field theoretical strings implies that the NG approximation is able to accurately reproduce their dynamics far beyond its naïve regime of validity. 
A direct consequence is that phenomena such as cusp formation - which arises rather generically in solutions of the Nambu--Goto equations for cosmic string loops \cite{Turok:1984cn} - should not be obstructed by higher-order corrections, and therefore is expected to occur in the full field theory evolution of these defects, as has been demonstrated in earlier simulations \cite{Olum:1998ag}.

Furthermore, we have developed a systematic procedure to include the possibility of excitation of bound states in the effective action. The amplitudes of such states play the role of massive scalar fields on the worldsheet that couple non minimally to the worldsheet curvature. We have identified the relevant couplings from a top down approach, and confirmed our predictions against numerical simulations of the string position. We have shown that, just including the lowest energy bound state in the effective action allows to reproduce the behavior of the string position with astounding accuracy through the whole regime of validity of the effective action.

Our results imply that, despite the lack of curvature correction terms at tree level in the effective action, the worldsheet gets curvature corrections due to the \emph{backreaction} of the excited bound states. In the language of QFT, these would be interpreted as loop corrections, and effective higher curvature correction could be generated by properly integrating out the massive modes from the effective action, resulting in a more `coarse-grained' effective action that involves only the Goldstone field. However, such effective action would struggle to describe some important physical processes that involve energy transfer mechanisms between the Goldstone and bound states, such as the parametric instability on the excited string  that we have numerically identified in this work.

On the other hand, our effective action including a bound state correctly predicts this instability, which is explained by the lowest order interaction between the Goldstone and shape modes through the linear coupling to the worldsheet Ricci scalar, which acts as a source. We remark that such coupling is not exclusive to this particular model, but a generic consequence of the geometric construction of the effective action. Hence, we conjecture that a similar instability mechanism will be present in the dynamics of extended defects that are excited with a bound state. This is in agreement with previous similar studies of excited domain walls in various dimensions \cite{Blanco-Pillado:2022rad, Blanco-Pillado:2024bev} and the axionic (global) string \cite{Blanco-Pillado:2022axf}. 
More broadly, this instability may serve as a diagnostic for identifying a significant population of bound excitations in field theory solitonic configurations, for example, following the numerical experiments performed in \cite{Blanco-Pillado:2020smt,Blanco-Pillado:2022rad}. This, in turn, could enable numerical simulations to more directly assess the potential cosmological relevance of these modes in cosmic string networks, as proposed in \cite{Hindmarsh:2021mnl}.

Finally, we note that the methods and techniques developed in this work can be straightforwardly extended to incorporate additional bound state excitations in the effective action, and potentially to determine their couplings to radiative degrees of freedom.

\acknowledgments
We are grateful to Daniel Jiménez-Aguilar, Ken D. Olum and Andrzej Wereszczynski for stimulating discussions and comments on previous versions of the manuscript. 
We would also like to thank Katy Clough, Thomas Helfer, Eugene Lim, Dina Traykova and the GRTL collaboration (\url{www.grtlcollaboration.org})
for their support and code development work. This work has been supported in part by the PID2021-123703NB-C21 and PID2024-156016NB-I00 grants funded by MCIN/AEI/10.13039/501100011033/and by ERDF;“ A way of making Europe”; the Basque Government grant (IT-1628-22) and the Basque Foundation for Science (IKERBASQUE). The work of AGMC is also supported by Grants No. ED481B-2025/059 and ED431B-2024/42 (
Xunta de Galicia). JCA acknowledges funding from the Department of Physics at MIT through a CTP Postdoctoral Fellowship. JQ has been supported in part by Spanish Ministerio
de Ciencia e Innovaci\'on (MCIN) with funding from the
European Union NextGenerationEU (PRTRC17.I1) and
the Consejer\'ia de Educaci\'on, Junta de Castilla y Le\'on,
through QCAYLE project, as well as the grant
PID2023-148409NB-I00 MTM, and the project Programa C2 from the University of Salamanca.

This work used Stampede3 at the Texas Advanced Computing Center through allocation PHY250054 from the Advanced Cyberinfrastructure Coordination Ecosystem: Services \& Support (ACCESS) program \cite{ACCESS}, which is supported by U.S. National Science Foundation grants 2138259, 2138286, 2138307, 2137603, and 2138296. The authors acknowledge the Texas Advanced Computing Center (TACC) at The University of Texas at Austin for providing computational resources that have contributed to the research results reported within this paper. URL: \url{www.tacc.utexas.edu}.


\appendix
\section{Review of higher codimension brane geometry}
\label{geo_review}
Consider a $p$-dimensional brane worldvolume in a $D$-dimensional spacetime. The worldvolume can be embedded as a $p$-dimensional hypersurface parametrized by $X^\mu(\sigma^A)$. 
We define a set of \emph{adapted coordinates} as the chart ${\xi^\mu}$ defined by:
\begin{equation}
    y^\mu(\xi)=X^\mu(\sigma^A)+n_i^\mu u^i,
\end{equation}
where 
\begin{equation}
    \xi^\mu=\left\{\begin{array}{cc}
        \sigma^A, &\mu\equiv A=0,1,\cdots p  \\
         u^i, &\mu\equiv i=1,\cdots D- p,
    \end{array}\right.
\end{equation}
and we have defined the tangent and normal vectors to the worldvolume as:
\begin{equation}
    e^\mu_ A=\pdv{X^\mu}{\sigma^A},\qquad g_{\mu\nu}n^\mu_ ie^\nu_A=0,
\end{equation}
and the associated dual forms satisfy:
\begin{equation}
    n_i ^\mu n_\mu ^j=\delta^j_i,\quad e^\mu_A e_\mu^B=\delta_A^B,\quad e^\mu_A e^A_\nu+n_i ^\mu n_\nu ^i=\delta^\mu_\nu.
\end{equation}
We will also be interested in the derivatives of a vector in the directions parallel to the worldvolume (i.e. the projection of the covariant derivative of a vector along the tangent bundle of the worldvolume submanifold). It is enough to study such projection for the tangent and normal basis vectors:
\begin{align}
    e^\mu_A\nabla_\mu e^\nu_B=\Gamma^C_{AB}e^\nu_C-K^i_{AB}n_i^\nu,\\[2mm]
    e^\mu_A\nabla_\mu n^\nu_i=\beta^j_{Ai}n^\nu_j+K^{iB}_{\,A}e_B^\nu,
    \label{derivs_en}
\end{align}
where $\Gamma_{AB}^C$ is the Levi-Civita connection of the induced metric on the worldvolume hypersurface, given by
\begin{equation}
    \gamma_{AB}=g_{\mu\nu}e^\mu_A e^\nu_B,
    \label{inducedmetric}
\end{equation}
 while $K_{AB}^i$ is its extrinsic curvature tensor:
\begin{equation}
    K_{AB}^i=-n^i_\mu e^\alpha_A\nabla_\alpha e^\mu_B,
\end{equation}
and $\beta^i_{jA}$ is the \emph{twist connection},
\begin{equation}
    \beta_{iA}^j=n^j_\mu e^\nu_A\nabla_\nu n^\mu_ i
\end{equation}
i.e. the metric connection on the normal bundle. Metric compatibility implies anti-symmetry under permutations of its normal indices, which allows to write the twist connection in terms of a twist vector $\omega_ A$ on the worldvolume:
\begin{equation}
    \eta_{ij}\beta^j_{kA}=\beta_{ikA}=-\beta_{kiA}\implies \beta_{ij A}=\epsilon_{ij}\omega_ A.
\end{equation}

In the new coordinate system, the spacetime metric can be written:
\begin{equation}
    g_{\mu\nu}\to g_{\alpha\beta}=\pdv{y^\mu}{\xi^\alpha}\pdv{y^\nu}{\xi^\beta}g_{\mu\nu}\equiv t_\alpha^\mu t_\beta^\nu g_{\mu\nu}
\end{equation}
where 
\begin{align}
    t_A^\mu=e^\mu_ A+u^ie^\nu_A\nabla_\nu n_i^\mu&= e^\mu_B(\delta_A^B+u^iK_{i\,A}^{\,\,B})+u^i\beta_{Ai}^jn_j^\mu,\notag\\[2mm] t_i^\mu&=n_i^\mu.
\end{align}
Therefore, we have the following splitting of the spacetime metric:
\begin{equation}
    g_{\alpha\beta}=\mqty(g_{AB}&g_{iB}\\g_{Aj}&g_{ij})
\end{equation}
where
\begin{align}
    g_{AB}=\gamma_{AB}+&2K_{jAB}u^j+K^{C}_{jA}K_{kBC}u^{j}u^k+\omega_ A\omega_B\delta_{ij}u^{i}u^j\notag\\
    g_{Aj}&=\epsilon_{ij}\omega_Au^i,\qquad g_{ij}=\delta_{ij}.
    \label{metric}
\end{align}
The metric determinant can be computed first by noting that, being a gauge-invariant quantity, must not depend on the (gauge dependent) twist connection, so we may as well use the gauge $\omega_A=0$. Then,
\begin{equation}
    \det(g_{AB})=\det(\gamma_{BC})\det(\delta_A^C+H_A^C),
\end{equation}
where $H_A^B$ is the $2\times 2$ matrix:
\begin{equation}
H^B_A=2K_{jA}^Bu^j+K^{C}_{jA}K_{kC}^Bu^{j}u^k
\end{equation}
and therefore we can write
\begin{align}
    \Tr H =2\bar{K}_1+\bar{K}_2,\\
    \Tr(H^2)=4\bar{K}_2+4\bar{K}_3+\bar{K}_4,\notag
\end{align}
where $\bar{K}_n\equiv \Tr{(u^i\bm{K}_{i})^n}$, i.e.
\begin{equation}
    \bar{K}_1=u^i \tensor{K}{_{i}^A_A}, \qquad \bar{K}_2= u^i u^j\tensor{K}{_{i}^A_B}\tensor{K}{_{j}^B_A}.
    \label{defK_K2}
\end{equation}
Further, the Cayley-Hamilton theorem for a $2\times 2$ matrix $\bm{A}$ yields
\begin{equation}
    \bm{A}^2=\bm{A}\Tr \bm{A}-\det(\bm{A})\bm{1}
\end{equation}
multiplying by $\bm{A}^{n-2}$ and taking the trace, we obtain a recurrence formula relating traces of higher powers of the matrix to the trace and determinant of the original matrix,
\begin{equation}
    \Tr( \bm{A}^n)=\Tr \bm{A} \Tr (\bm{A}^{n-1})-\det(\bm{A})\Tr (\bm{A}^{n-2} ).
    \label{recurrel}
\end{equation}
Applying \eqref{recurrel}, we may write $\det(\delta_A^B+H_A^B)$ in terms of $\Tr \bm{H}$ and $\Tr (\bm{H}^2)$, and therefore in terms of $\bar{K}_{1,2,3,4}$. Eqn. \eqref{recurrel} again relates $\bar{K}_n$ with $\bar{K}_1$ and $\bar{K}_2$ for $n\geq 3$. Finally, rearranging terms we arrive to 
\begin{align}
   \det(\delta_A^B+H_A^B)=&\qty[1+\bar{K}_1+\frac{1}{2}(\bar{K}_1^2-\bar{K}_2)]^2 =\notag\\
   =&[\det(\delta_A^B+u^iK_{i\,A}^{\,\,B})]^2
\end{align}
and Eqn. \eqref{det_effmet} of the main text follows. We remark that the condition

\begin{equation}
    \det(\delta_A^B+u^iK_{i\,A}^{\,\,B})=0
\end{equation}
determines the domain of validity of such adapted coordinate system.
\section{Derivations of the effective action terms}
\label{app:Derivations}
A simple way of computing the coupling terms between the worldsheet geometry and the scalar field in the effective action \eqref{dS1} is to compute the functional variation of the action in flat space coordinates and then transform the result into adapted coordinates.

\subsection{Tadpole term}
Let us consider first the tadpole term in the effective action. the first variation of the kinetic term for the scalar field is given by
\begin{align}
    \frac{\delta S}{\delta\phi}\Bigg\rvert_{\bar \phi}^{(\rm kin)}=&\eta^{\alpha\beta}\partial_{\alpha}\partial_\beta\phi\Bigg\rvert_{\bar \phi}=\eta^{\alpha\beta}t^\mu_\alpha\nabla_\mu\qty(t^\nu_\beta\nabla_\nu \bar\phi).
\end{align}
Now, $\bar{\phi}$ depends only on spacelike adapted coordinates. Thus, 
$t^\nu_\beta\nabla_\nu \bar\phi=n_\beta^i\nabla_i\bar\phi$. Also,
\begin{equation}
    \nabla_\alpha n^i_\beta=\nabla_An_\beta^i=e_A^\lambda \partial_\lambda n_\beta^i=\beta^i_{jA}n_\beta^j+K^{i}_{AB}e_\beta^B
\end{equation}
where we have used \eqref{derivs_en} in the last step. On the other hand, since the terms in the effective action must also be gauge independent, we may use the gauge $\beta_{jA}^i=0$ in the following derivations. Then,
\begin{align}
\eta^{\alpha\beta}\partial_{\alpha}\partial_\beta\phi\Big\rvert_{\bar \phi}  =&\eta^{\alpha\beta} \qty(t_\alpha^AK_{AB}^ie_\beta^B\nabla_i \bar\phi +t_\alpha^it_\beta^j\nabla_{ij}\bar\phi)\label{kinetic_term_phi}\\
=&\eta^{\alpha\beta}(\delta_A^C+u^lK_{l\,A}^{\,\,C})^{-1}e^C_\alpha K^i_{AB} e^B_\beta\nabla_i \bar \phi+\nabla^2\bar\phi\notag.
\end{align}
The first term in the rhs of \eqref{kinetic_term_phi} is precisely the scalar part of \eqref{dS_BPS}. The second term is part of the scalar field static equation of motion in the adapted coordinate system. The rest of the terms come from the variation of the potential part, which does not depend on the coordinate system. Therefore, this term will cancel when evaluated on the BPS solution. 

For the vector part of $\delta^{(1)}S$, we follow the same reasoning. We will need only consider the kinetic part,
\begin{equation}
    \frac{\delta S}{\delta A^\mu}\Bigg\rvert_{\bar A}^{(\rm kin)}=\eta^{\alpha\beta}\partial_\alpha  F_{\beta \mu}\Big\rvert_{\bar A}=\eta^{\alpha\beta}\partial_\alpha\qty(t^\lambda_\beta t^\gamma_\mu\bar F_{\lambda\gamma}).
\end{equation}
Again, the BPS solution only depends on the spacelike adapted coordinates. Therefore,
${t^\lambda_\beta t^\gamma_\mu\bar F_{\lambda\gamma}=n^i_\beta n^j_\mu\bar F_{ij}}$, and 
\begin{align}
\eta^{\alpha\beta}\partial_\alpha(n^i_\beta n^j_\mu\bar F_{ij})&=\eta^{\alpha\beta}\qty(t^A_\alpha\nabla_A n^i_{(\beta} n^j_{\mu)}\bar F_{ij}+n^i_\beta n^j_\mu n^k_\alpha\nabla_k\bar F_{ij})\notag\\
&=t^A_\alpha K_{A}^{i\,\, B}e_B^\alpha n^j_{\mu}\bar F_{ij}+ n^j_\mu \nabla^k\bar F_{kj}.
\end{align}
As in the scalar case, the last term in the rhs corresponds to (part of) the static equation of motion, and will vanish on shell once the rest of the terms are taken into account. The first term, however, yields the vector part of \eqref{dS1} when contracted with $a^\mu$.

\paragraph*{Integrating along codimensions}{The expression \eqref{S_expansion} is not yet in the desired form as it still depends on the codimension coordinates. We will now explain the procedure to integrate out along the codimension, so that it can be rewritten as an effective linear coupling terms between worldsheet geometric invariants and the scalar field that represents the amplitude of the excited bound state. Such terms are constructed out of the traces of products of the extrinsic curvatures, $K^{iA}_{\,\, B}$, and must be both invariant under worldsheet diffeomorphisms and under $SO(2)$ rotations on the normal bundle indices. Furthermore, as argued, for a $2$-dimensional worldsheet the extrinsic curvatures are $2\times 2$ matrices, so any invariant involving powers of the curvatures reduces to $SO(2)$-symmetric combinations of the following invariants
\begin{equation}
    P^i=\tr(\bm{K}^i),\quad Q^{ij}=\Tr(\bm{K}^i\bm{K}^j).
    \label{invariants}
\end{equation}
Furthermore, the coupling coefficients can also be computed systematically, as we will now explicitly show.

We start by noting that the trace in Eqn. \eqref{dS_BPS} can be expressed in a series of (traces of) powers of $u_iK^{iA}_{\,\,B}$. Remarkably, the Cayley-Hamilton theorem ensures that such expansion truncates at a finite order, which corresponds to the dimension of the worldsheet. 
Indeed, we have
\begin{equation}
    (\bm{1}+u_l\bm{K}^l)^{-1}=\frac{(1+\bar K_ 1)\bm{1}-u_ i\mathbf{K}^i}{1+\bar{K}_1+\frac{1}{2}(\bar{K}_1^2-\bar{K}_2)},
    \label{Cayley-inverse}
\end{equation}
so the tadpole term in \eqref{S_expansion} can be written
\begin{equation}
    \delta^{(1)}S=\int\!\!\int \qty[(1+u_jP^j)P^i-u_jQ^{ij}]\mathcal{F}_i \sqrt{-\gamma}\, d^2u d^2 \sigma
\end{equation}
where 
\begin{equation}
    \mathcal{F}_i= \nabla_ i\bar{\phi}\varphi+\eta^{jk}\bar{F}_{ij}a_k
\end{equation}
Using polar coordinates in the normal space, we arrive at the effective term of Eqn. \eqref{coupling_R}. Remarkably, the extrinsic invariants \eqref{invariants} appear in a specific combination, such that the final result can be written exclusively in terms of intrinsic geometry of the worldsheet, namely, the Ricci scalar.
\subsection{Quadratic term}
Let us now focus on the quadratic terms in the worldsheet effective action. the second order variation can be written
\begin{equation}
    S^{(2)}=-\frac{1}{2}\int d^4x \Psi^T\mathcal{H}\Psi
\end{equation}
where $\Psi\equiv (\delta\phi,\delta A_\mu)$, and $\mathcal{H}$ is the associated Hessian operator. Introducing adapted coordinates, such operator separates as
\begin{equation}
    \mathcal{H}=\mathcal{H}_\parallel(u^i,\nabla_ A)\oplus\mathcal{H}_\perp( u_ i,\partial_ i)
\end{equation}
where $\mathcal{H}_\parallel$ involves worldvolume derivatives, but $\mathcal{H}_\perp$ only depends on transverse coordinates and their derivatives. This justifies the ansatz \eqref{ansatz_perturb} for the perturbations in terms of the shape mode, $\Phi_{\rm shape}(u)=(\varphi(u),a_\nu(u))$, which is a discrete normal mode of $\mathcal{H}_\perp$ with positive eigenvalue $\omega^2$, localized along the transverse direction, as explained in \Cref{Sec:Seff+BS} of the main text.
Indeed, let
\begin{equation}   \Psi(\sigma,u^i)=\chi(\sigma)\Phi_{\rm shape}(u)
\end{equation}
such that 
\begin{equation}
    \mathcal{H}_\perp \Phi_{\rm shape}(u)\equiv \mathcal{L }\Phi_{\rm shape}(\rho)=\omega^2 \Phi_{\rm shape},
\end{equation}
then 
\begin{equation}
    \mathcal{H}\Psi=\qty(-g^{AB}\nabla_{AB}+\omega^2)\chi(\sigma^A)\Phi_{\rm shape}(u^i)
\end{equation}
and we end up with
\begin{equation}
    S^{(2)}=-\frac{1}{2}\int \chi\qty(-g^{AB}\nabla_{AB}+\omega^2)\chi|\Phi_{\rm shape}|^2\sqrt{-g}d^2\sigma d^2 u.
    \label{effective_quadratic}
\end{equation}
Written in this way, the integration along transverse directions is still nontrivial, due to the dependence of the metric on the transverse coordinates. Ideally, we would be able to expand \eqref{effective_quadratic} in a finite series of terms, ordered in powers of the transverse coordinates. This was indeed possible for the metric determinant (Eqn. \eqref{det_effmet}), and also for the tadpole term. However, we will now see that this is no longer the case for the inverse metric $g^{AB}$, and we will have to consider an infinite series instead.

Indeed, starting from the defining expression of the worldvolume metric \eqref{metric}, it is straightforward to show that (in the $\omega_A=0$ gauge):
\begin{equation}
    g_{AB}=(\gamma_{AC}+u^iK_{i AC})\gamma^{CD}(\gamma_{DB}+u^iK_{i DB}).
\end{equation}
Hence, using the relation \eqref{Cayley-inverse}, we get 
\begin{equation}
    g^{AB}=\frac{(1+\bar K_ 1)\gamma^{AB}-2 u^l K_ l^{AB}(1+\bar K_ 1)+u^l u^k K_{l\, D}^A K_{k}^{DB}}{[1+\bar{K}_1+\frac{1}{2}(\bar{K}_1^2-\bar{K}_2)]^2}.
    \label{g_inv}
\end{equation}
The presence of the inverse determinant in  \eqref{g_inv} makes it impossible for the expansion in terms of powers of $u^i$ to truncate at any finite order. For the lowest nontrivial order, we find,
\begin{align}
    &\sqrt{-g} g^{AB}=\sqrt{-\gamma}\times\big[\gamma^{AB}-2 u^lK_{l}^{AB}\notag\\
    &+\frac{1}{2}(\bar{K}_1^2+\bar K_2)\gamma^{AB}+u^l u^k K_{l\, D}^A K_{k}^{DB}+\order{u^3}]
\end{align}
We might then integrate along the codimensions. Using again polar coordinates and taking into account the symmetry properties of $\Phi_{\rm shape}$, we get 

\begin{equation}
    S^{(2)}_{\rm eff}\approx -\frac{1}{2}\int \chi\qty[-g^{AB}_{\rm eff}\nabla_{AB}+\omega^2(1+c_2\mathcal{R})]\chi\sqrt{-\gamma} d^2\sigma+\cdots
\end{equation}
where 
\begin{equation}
    g^{AB}_{\rm eff}=\qty[1+c_2\qty(\frac{1}{2}\mathcal{R}+(K_{lA}^{\, A})^2)]\gamma^{AB}+ c_2 K^{A}_{lC}K_{l}^{CB}
\end{equation}
and 
\begin{equation}
    c_2=2\pi\int_0^\infty \rho^3 |\Phi_{\rm shape}|^2 d\rho\approx 2.59\pi .
    \label{c2}
\end{equation}
Therefore, the effective action for the 
shape mode amplitude is nonlocal, and can be written as an infinite series in powers of the worldsheet curvature radius. The expansion can be truncated at any order and generates an interacting field theory for a scalar field that sees an ``effective metric'' and which is coupled to the worldsheet geometry in a non trivial way. The price to pay for such truncation is the appearance of  higher time derivatives in the equations of motion for the massive field and the Goldstone modes, unless the curvature expansion of the effective metric is truncated at zeroth order, which yields Eqn. \eqref{dS2_eff}.
\section{Effective action for the collapsing loop}
\label{app:circular}
As an example of the action developed in Section \ref{Sec:Seff+BS}, we present the explicit effective action for a collapsing loop with one massive degree of freedom. We choose the following coordinates for the string position:
\begin{equation}\label{loop_coord}
X^\mu=\left(\tau, R \cos\theta, R\sin\theta,0\right)
\end{equation}
The intrinsic metric has the following form
\begin{equation}
\gamma=\left(
\begin{matrix}
1-\dot R^2 & 0\\
0 &-\dot{R}^2
\end{matrix}\right)
\end{equation}
and the Ricci scalar is
\begin{equation}
\mathcal{R}=\frac{2\ddot{R}}{R\left(1-\dot{R}^2\right)^2}.
\end{equation}
Taking into account (\ref{ansatz_perturb}), the action (\ref{S_expansion}), in coordinates (\ref{loop_coord}), takes the following form
\begin{eqnarray}
S &=& - \int d\tau \, R \sqrt{\dot{R}^2 - 1} \Big(
\pi - \frac{1}{2(1-\dot{R}^2)} \dot{\chi}^2 + \frac{1}{2} \omega^2 \chi^2 \nonumber\\
&& \hspace{2em} + \kappa \chi \mathcal{R}+\frac{1}{2}\omega^2c_2 \chi^2 \mathcal{R}+\mathcal{O}\left(\chi^3\right) \Big)
\end{eqnarray}
The equations derived from this action are not particularly illuminating; however, it is important to emphasize that, since $\mathcal{R}$ is topological in $1+1$ dimensions, the terms proportional to $\kappa$ and $c_2$ introduce at most second-order time derivatives.

\section{Convergence testing}

We study the convergence of our numerical simulations by repeating the run shown in
Fig.~\ref{fig:loop_2d_panel} at three base resolutions: low ($N=32^3$), medium ($N=64^3$), and high ($N=128^3$), with an extra level of mesh refinement. For each resolution we extract a lineout of $\vert\phi\vert^2$ along the $y$-axis in the plane $z=0$ at $t\approx 550$. This corresponds to profiles near the string core and at the approximate time when the zero-mode amplitude is maximum. We define the error in the profiles $\vert\phi\vert^2$ between two resolutions as
\begin{equation}
    \Delta \left[|\phi|^2\right] \equiv \mathrm{abs}\left(|\phi|^2_{\mathrm{res}_1} - |\phi|^2_{\mathrm{res}_2}\right) .
\end{equation}
Figure \ref{fig:convergence} plots $\Delta\left[\vert\phi\vert^2\right]$ for resolution pairs and  shows convergence between 2nd and 4th order as the resolution is increased.
\begin{figure}[b]
    \includegraphics[width=\linewidth]{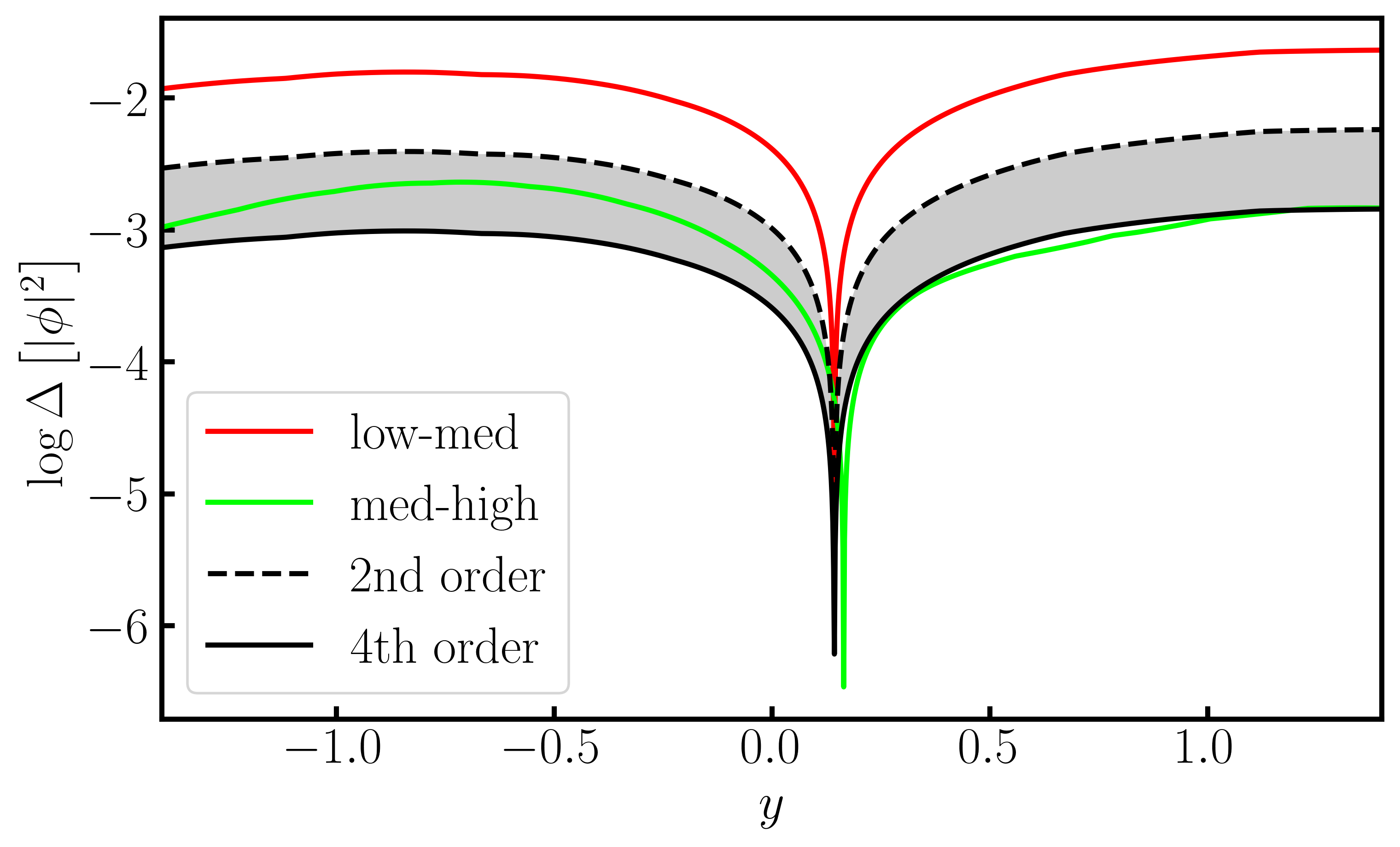}
    \caption{Convergence test of the parametric instability. We extract the $\vert\phi\vert^2$ profile along the $y$-axis at $z=0$ and $t\approx 550$. The plot compares simulations with base resolutions $N_\mathrm{low}=32^3$, $N_\mathrm{med}=64^3$ and $N_\mathrm{high}=128^3$, all with one level of refinement. The figure shows between 2nd and 4th order convergence of the error.}
    \label{fig:convergence}
\end{figure}

\bibliography{Bibliography}
\end{document}